\documentclass[%
 rsi,
amsmath,amssymb,
reprint,%
superscriptaddress,
]{revtex4-1}

\usepackage{graphicx}
\usepackage{dcolumn}
\usepackage{bm}
\usepackage{xcolor}

\usepackage[utf8]{inputenc}
\usepackage[T1]{fontenc}
\usepackage{mathptmx}
\usepackage{booktabs}
\usepackage{exscale}
\usepackage{relsize} 

\begin{document}

\preprint{AIP/123-QED}

\title[On-site transverse position compensation]{Numerical study of transverse position monitor and compensation \\ for x-ray polarization diagnosis}

\author{Zipeng Liu}
\affiliation{Shanghai Institute of Applied Physics, Chinese Academy of Sciences, China}
\affiliation{University of Chinese Academy of Sciences, China}

\author{Bangjie Deng}%
\affiliation{Xi'an Jiaotong University, School of Nuclear Science and Technology, China}%

\author{Haixiao Deng}
\email{denghaixiao@zjlab.org.cn}
\affiliation{Shanghai Advanced Research Institute, Chinese Academy of Sciences, China}

\author{Bo Liu}
\affiliation{Shanghai Advanced Research Institute, Chinese Academy of Sciences, China}

\date{\today}

\begin{abstract}
        Diagnosing free electron laser (FEL) polarization  is critical for polarization-modulated research such as x-ray free electron laser (XFEL) diffraction imaging and probing material magnetism. In an electron time-of-flight (eTOF)\ polarimeter, the flight time and angular distribution of photoelectrons were designed  based on x-ray polarimetry  for on-site diagnosis. However, the transverse position of x-ray FEL pulses introduces error into the measured photoelectron angular distribution. This work thus proposes a method to monitor  the transverse position using an eTOF polarimeter and explains how to compensate for the error due to transverse position. A comprehensive numerical model is developed to demonstrate the feasibility of the compensation method, and the results reveal that a spatial resolution of 20 \(\mu\)m and a polarity improved by 0.5\% is possible with fully polarized FEL pulses. The impact of FEL\ pulses and a method to calibrate their linearity  is also discussed.
\end{abstract}

\maketitle

\section{Introduction}

Being fully coherent and offering controlled polarization  \cite{Allaria2014,Deng2014,Lutman2016,Huang2020},
the x-ray free electron laser (XFEL) is at the heart of  advanced  techniques in fields such as materials science and surface physics. 
Currently, several XFELs machines have been constructed or are in operation \cite{Ackermann2007,Emma2010,Ishikawa2012,Feng2018}. 
The diagnosis of XFEL pulse polarization is critical in research involving  polarization-modulated diffraction and spectroscopy  \cite{Schtz1987,Chen1990,Kortright1999,2005Polarization}.
Several  methods are available to measure the x-ray polarization of XFEL pulses, including phase-shift retarders \cite{Gaupp1989,Koide1991,Hochst1994,Schafers1999,Vodungbo2011} and angular-distribution-based  polarimeters \cite{Lutman2016,Laksman2019}. 
In particular, the time of flight and angular distribution of photoelectrons produced by x-ray polarimeters are used for on-site diagnosis, in so-called 'eTOF polarimeters'. 
Such instruments have been used in main FEL facilities \cite{Zhang2017Design,2019Commissioning,2021Multi}. 
The Shanghai soft-x-ray FEL facility \cite{2021NanshunFeatures} is currently offline testing an eTOF-like prototype for angular-resolved polarimeter (ARPolar). 
However, variations in the incident position of XFEL pulse may distort the photoelectron's angular distribution and thus invalidate the resulting polarization diagnosis.
 
A single-shot x-ray beam-position monitor (XBPM) is essential for XFEL diagnosis and experimentation. Various XBPMs have been developed in FEL facilities, such as  fluorescence detectors \cite{Alkire2010Design,Heimann:yi5066}, ion chambers \cite{Ilinski2007Residual}, and back-scattering photodiodes \cite{Tono2013}. However,  the low intensity of  the soft XFEL makes it  difficult to use the fluorescence screen and backscattering methods to determine beam position. Although another ion chamber nearer  the eTOF polarimeter may be used to obtain the transverse position, it is hard to rearrange the machine. Therefore,  the shot-to-shot transverse position must be measured so that the error term may be compensated. 
 
 Since monitoring transverse position and  polarity  are both indispensable for XFEL experiments, the ARPolar device must offer these two functions.
 This work thus proposes an on-site single-shot x-ray beam position monitoring and polarization compensation method using the ARPolar polarimeter. The methods and principles are first described, following which  a comprehensive numerical model is established to validate the method, and the calibration method is presented. Finally, the key factors are discussed, including detector linearity  and  photoelectron yield.
 
\section{Principles and methods}

\begin{figure*}
    \centering
    \includegraphics[width=\linewidth]{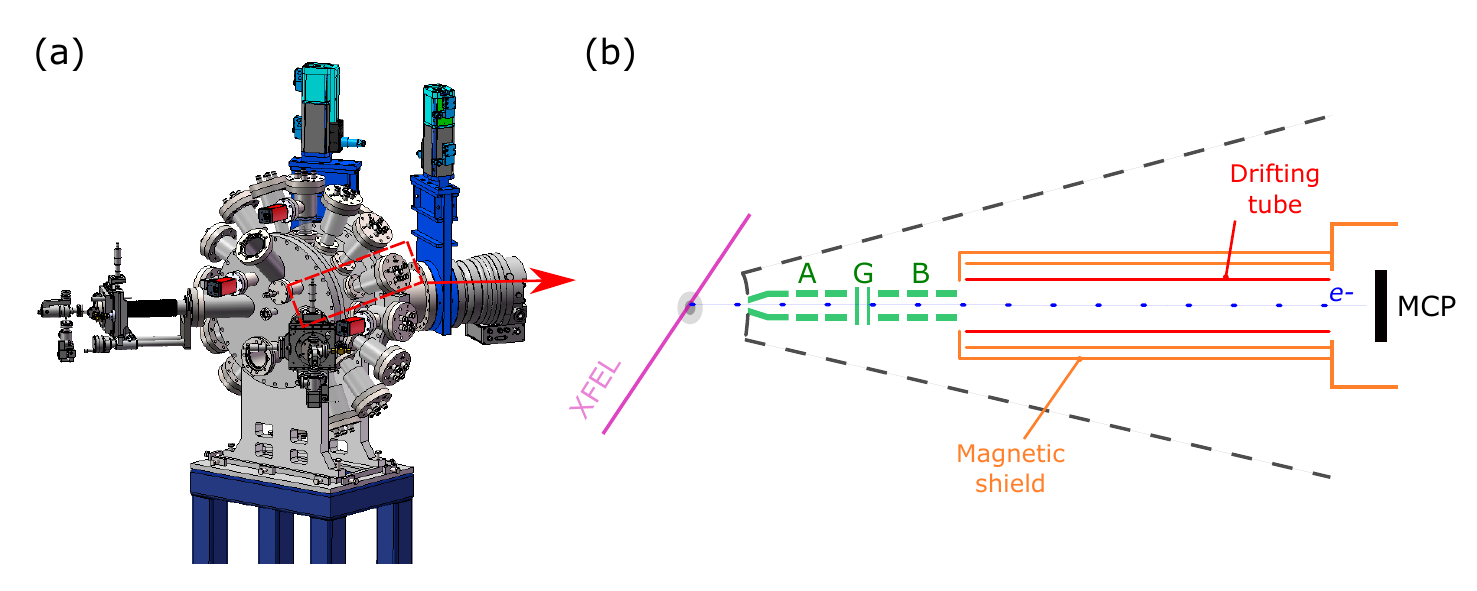}
    \caption{The ARPolar instrument: (a) drawing of  instrument,  (b)  scheme of a detection channel, positioned in  red square in panel  (a). Detection channel  consists of focusing electrode groups (A and B), retarding grid electrodes (G), magnetic shields, a drift tube, and a MCP detector. }
    \label{fig:arpolar instrument}
\end{figure*}

 Figure \ref{fig:arpolar instrument} shows the ARPolar instrument, which consists of a chamber, 16 detection channels, two gas-injection assemblies, and two turbo pumps.
Each channel contains three groups of electrodes, a magnetism-shielded flight tube, and a microchannel plate detector (MCP).
Electrons produced by XFEL pulses ionize a gas target in the center of the chamber. The electrons are dipole-symmetrical in the plane perpendicular to the beam-line (detection plane), from which the polarization characteristics  of the XFEL\ pulses can be obtained. The angular distribution can be divided into four identical ranges, starting from the polarization angle and then incrementing by \(\pi/2\).
However, this symmetry in the angular distribution changes when the x-ray pulses are displaced from the center of the beamline pipe.
By solving for the incident positions, both the real detection direction and the correct charge distribution can be predicted. Furthermore, the linear polarity can be compensated accordingly. The corresponding principles and the methods are presented here.

\begin{figure}[tbhp]
    \centering
        \includegraphics[width=0.8\linewidth]{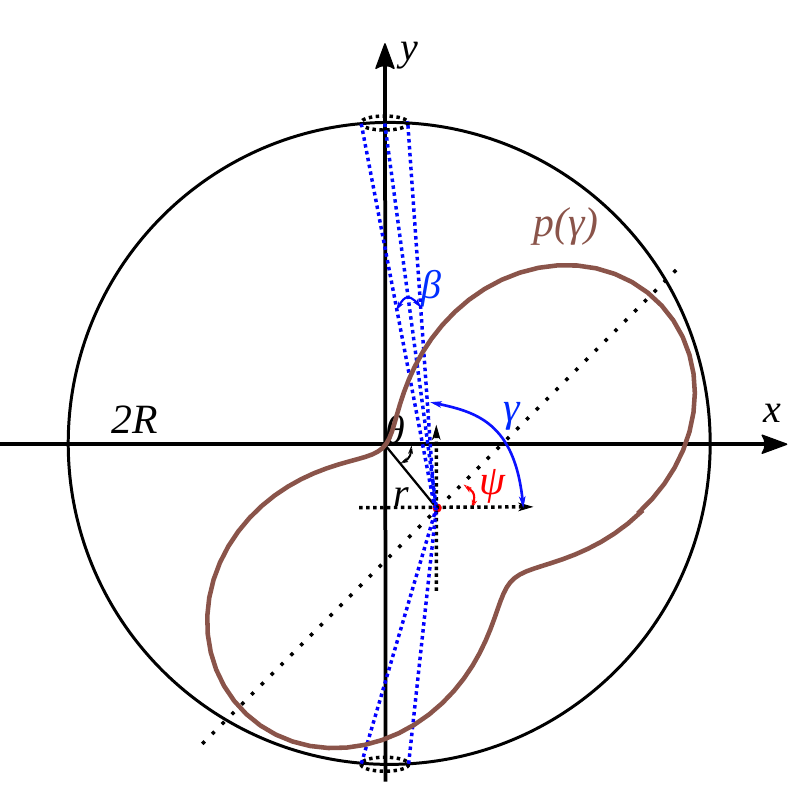}
        \caption{Diagram of  detection angle.}
        \label{fig:diagram of detection angle}
\end{figure}


 Considering the incident position \(P(r,\theta)\), the charge  \(Q(r,\theta)\) detected in the detection plane  can be written as 
 \begin{align}
     Q(r,\theta)=&\int \frac{p(\gamma)}{\rho^2}{d}\gamma , \label{eq:polar coordinate system angular distribution of photon-electron} \\
      \gamma(r,\theta)=& \mathrm{arccos}\left(\frac{R\cos\gamma-\gamma\cos\theta}{\sqrt{R^2+r^2-Rr\sin(\theta-\gamma)}}\right), \label{eq:gamma angle}
 \end{align}
 where \(p(\gamma)\) is the angular distribution of electrons, \(\rho\) is the distance from the pulse center to the detection hole of the channel, \(r\) is the relative displacement between the device center  and the x-ray pulse, \(R\) is the distance from the front tube to the device center, \(\theta\) is the angular direction  of the incident position, and \(\gamma\) is the angular direction  of the detector (see  Fig. \ref{fig:diagram of detection angle}).
 The horizontal and vertical axes divide the electrons into four quadrants,  
 \begin{equation}  \label{eq:Qi charge}
    Q_i=\int_{\psi+\pi(i-1)/2}^{\psi+\pi i/2}{\frac{p(\gamma)}{\rho^2}} {d}\gamma,\quad  {i=1,...,4,} 
 \end{equation}
where \(\psi\) is the polarization angle of the x-ray pulse, which is determined by the undulator configuration. When \(\psi=0\), the position sensitivities \(S_x\) and \(S_y\), which are the proportionality constants between charge  and pulse displacement, are defined as \cite{Shafer2008,Gaupp1988}
\begin{align}
        S_x & = \dfrac{{d}}{dx}\left(\frac{\Delta Q_{x,\psi=0}}{\sum Q}\right) = \dfrac{{d}\eta_{x,\psi=0}}{{d}x} ,  \\
        S_y & =  \dfrac{{d}}{dy}\left(\frac{\Delta Q_{y,\psi=0}}{\sum Q}\right) = \dfrac{{d}\eta_{y,\psi=0}}{{d}y},
\end{align}
where \(S_x\) and \(S_y\) refer to the horizontal  and vertical directions, respectively. Furthermore,
\begin{align}
    \Delta Q_x &=(Q_1+Q_4)-(Q_2+Q_3), \\
    \Delta Q_y &=(Q_1+Q_2)-(Q_3+Q_4), \\
    \sum Q_x&=Q_1+Q_2+Q_3+Q_4 ,\\
    \eta_x &= \frac{\Delta Q_x}{\sum Q},\label{eq:eta_x} \\
    \eta_y &= \frac{\Delta Q_y}{\sum Q}.\label{eq:eta_y}
\end{align}

The relation \( x(\eta_x)\) can be expanded as a Taylor series about \(\eta_x = 0\):
\begin{equation}\label{eq:eta_x def}
        x(\eta_x) = x(\eta_x = 0) + x'(\eta_x)\eta_x + O(\eta_x),
\end{equation}
where \(\eta_x = 0\) indicates \(\Delta Q_x = 0\) and \(\hat{x} = 0\). Thus, the incident position (\(\hat{x}\), \(\hat{y}\))
\begin{align}
        \hat{x} &\approx \frac{1}{S_x}\eta_x + \delta_x, \label{eq:x predected}\\
        \hat{y} &\approx \frac{1}{S_y}\eta_y + \delta_y. \label{eq:y predected}
\end{align}
where \(\delta_x\) and \( \delta_y \) refer to the contribution of \(O(\eta_x)\) and \( O(\eta_y) \), respectively. 
In  \(x\text{-}y\) coordinates, the real position \((x,y)\) is obtained by rotating an angle \(\psi\), as follows:
\begin{equation}\label{local coordinate to global coordinate}
\left[
\begin{array}{l}
x \\
y
\end{array}
\right] = \left[
\begin{array}{cc}
\cos(\psi)&-\sin(\psi)\\
\sin(\psi)&\cos(\psi)
\end{array}
\right]
\left[
\begin{array}{l}
\hat{x} \\
\hat{y}
\end{array}
\right].
\end{equation}

 In the  vertical plane through the beamline, the angular distribution \( p_s(\gamma) \) of \(s\)-shell photoelectrons  can be described  in the dipole approximation as \cite{Manson1982,Cooper1993,Zhang2017Design} 
\begin{equation}\label{eq:Angular distribution of phot}
        p_s(\gamma)= A\{1+P_l\cos{[2(\gamma - \psi)]}\},
\end{equation}
where \(A\) is the normalization constant; \( P_l \) and \( \psi \) are  the linear polarity and the polarization angle of the FEL, respectively. The polarization characteristics of the XFEL can be obtained by fitting the angular charge distribution  monitored by detectors  to Eq. (\ref{eq:Angular distribution of phot}). For the incident position\((\hat{x}_i,\hat{y}_i)\). The average probability density \(\bar{p}\left(\gamma_i\right)\) is
\begin{equation}\label{eq:charge distribution at incident pos}
        \bar{p}_s(\gamma_i)=\frac{Q_{s,i}}{\beta_i}=\int_{\gamma_{i,-}}^{\gamma_{i,+}}p_s\left(\gamma\right)\mathrm{d}\gamma ,
\end{equation}
where \( Q_{s,i} \) is the number of charges collected by detector \( i \), \(\beta_i\) is the collection angle of detector \( i \), and \(\bar{p}\left(\gamma_i\right)\) is the average probability density for \(\beta_i\) and is formed using Eq. (\ref{eq:Angular distribution of phot}). 
The charges detected by detector \( i \) can then be found by using Eq. (\ref{eq:charge distribution at incident pos}). Fitting the  data to Eq. (\ref{eq:Angular distribution of phot}) gives the new linear polarity and polarization angle. 

\section{Numerical model}

To validate the method described above, we developed a comprehensive numerical model. A Monte Carlo  model was developed to  simulate the yield and  angular distribution of photoelectrons, whose primary photon source was configured according to a start-to-end simulation of the  FEL\ in self-amplified spontaneous emission (SASE) mode. The vertices of the photoelectrons and Auger electrons were recorded to simulate electron transmission, which was coupled with the static electric field  to obtain the flight trajectories and collection efficiency of the electron-optical system of the ARPolar soft-x-ray-FEL (SXFEL) polarimeter. In the following, the  model is described in detail.

\subsection{Model setup}

\begin{figure}[tb]
    \centering
        \includegraphics[width=\linewidth]{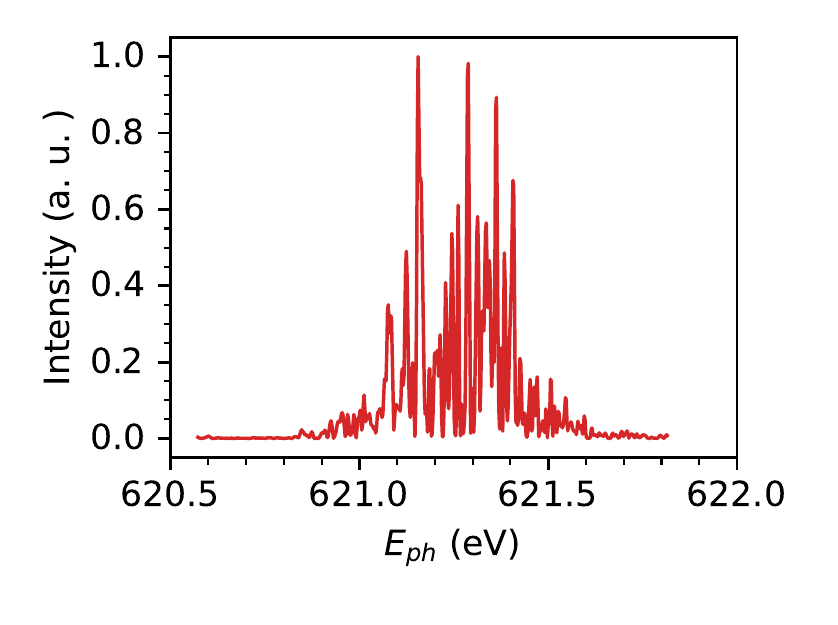}
        \caption{Energy spectrum of  FEL pulse.}
        \label{fig:Energy spectrum of the FEL pulse}
\end{figure}

The full process of the SASE-mode FEL  was simulated to produce FEL pulses based on the configuration of the SASE line of the Shanghai SXFEL, has an electron-beam energy of 1.6 GeV and an undulator period \(\lambda_u=16\) mm. The center wavelength \(\lambda_s\) of the XFEL pulse  can be determined from the  resonance condition:
\begin{equation}
        \lambda_s = \dfrac{\lambda_u}{2\gamma^2}\left(1+\dfrac{K^2}{2}\right),
\end{equation}
where \(\gamma\) is the electron-beam Lorentz factor and  \(K\) is the field parameter of the undulator. This process was simulated by the well-benchmarked FEL code GENESIS. The root-mean-square (RMS) value of the transverse diameter of the pulse is \(200\ \mu \text{m}\) and the energy spectrum of the FEL pulse is shown in  Fig. \ref{fig:Energy spectrum of the FEL pulse}. The simulation indicates the saturation power is on the order of \(10^9\) W and that each FEL pulse contains \(1.133\times10^{12}\)  photons, with an  average energy and  bandwidth  around 621  and 0.8 eV, respectively.


To simulate the yield and the angular distribution of the photoionized electrons, we developed a Monte Carlo model within the framework of the Geant4 code and based on our previous research \cite{Zhang2017Design}.
The geometry of the instrument was set up after reasonable simplifications. The  ARPolar polarimeter contained 16 detection channels, where, for each channel, the incident hole  was oriented  toward the center of the beamline pipe and captured  electrons traveling in a uniform angular direction. The vertices of the primary photons were located at the center of the chamber, and their energy was distributed as shown in  Fig. \ref{fig:Energy spectrum of the FEL pulse}. The ionization process instigated by the polarized photon was based on the  G4Livermore polarized electromagnetic model combined with the cross-section dataset G4EMLOW-4.8 \cite{1997EPDL97}. Table \ref{tab:Mente-Carlo parameters} summarizes the key parameters in the model.
\begin{table}[htp]
        \caption{Main parameters of the Geant4-based Monte Carlo model.}
        \label{tab:Mente-Carlo parameters}
        \begin{tabular}{ccp{4cm}}
                \toprule
                Parameter & Value & Description \\
                \midrule
                \(D\)         & 20 mm & Distance between the center of the beamline and the incident hole of a TOF channel. \\
                \(\phi_i\)  & 3 mm  & Diameter of the incident hole.                                                        \\
                \(\rho_{tg}\) & \(1.54\times10^{-11} \text{ kg/m}^3\) & Average gas density of the target. \\
                \(L_d\) & 400 mm & Drift length of photoelectrons. \\
                \(P_l\) & 1 & Linear polarity. \\
                \(\theta_p\) & 0 & Polarization angle. \\
                \bottomrule 
        \end{tabular}
\end{table}
As long as the electrons entering  the incident hole, the energies,  directions, and  positions of their primary vertex are recorded. The result indicates that 38\% of photo-ionized electrons generated within the detection plane are captured in the detection channels because the flight distance before entering the detection hole is only 20 mm and the chamber is well shielded against stray magnetic fields. For the full linearly polarized FEL pulses, about \(10^ 4\) incident electrons enter the channel oriented along the polarization direction, which suggests that the statistical fluctuation is sufficiently small  to diagnose polarization  and monitor position.

\begin{figure}[tb]
    \centering
    \includegraphics[width=\linewidth]{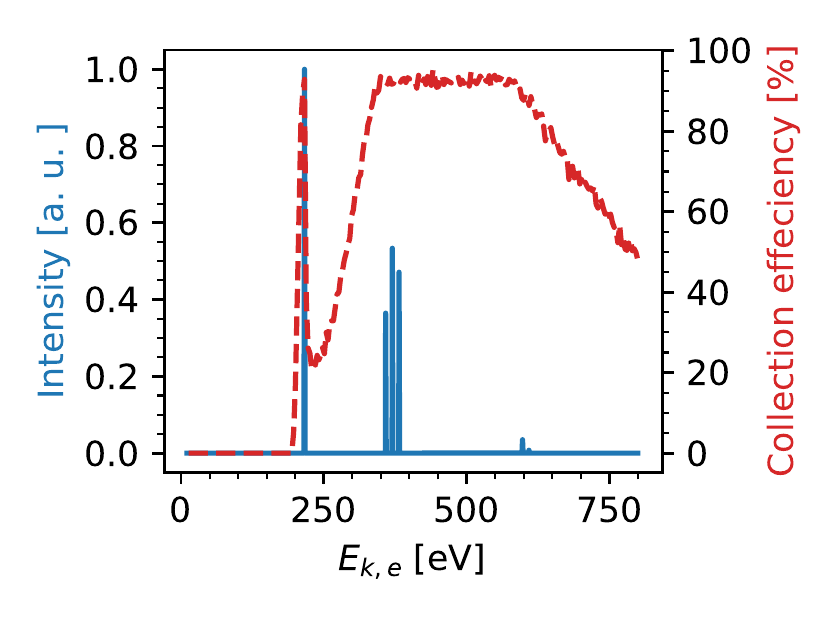}
    \caption{Collection efficiency of electron-optical system. }
    \label{fig:deteff}
\end{figure}

To simulate the collection efficiency of the electron-optical system, we developed a coupled static-electric,  particle-tracing  model  using the SIMION software \cite{1990SIMION}. The model consists of three electrode groups A, B, and G and a drift tube. Electrodes of groups A and B are three ring-type electrodes that focus and filter electrons. Group G electrodes are two grid electrodes \(\text{G}_1\) and \(\text{G}_2\) that retard  electrons for long  time intervals that depend on the electron energy. The  drift tube was 300 mm long and its  voltage was floated with respect to  ground and tuned to maintain the energy of the retarded electrons. The particle trajectories and the flight time were recorded to analyze the detection efficiency, which is defined as the ratio of the number of electrons collected  by the MCP detectors to the number of electrons incident on the detection hole. For a \(\text{N}_2\) target and FEL pulses with an average energy of around 621 eV, the \(s\)-shell electron energy is about 217 eV. we designed an  electron-optical system  to optimize the collection efficiency for photo-electrons and Auger electrons. Figure \ref{fig:deteff} shows the tuned collection efficiency, which shows the electron-optical system covers both the photo-electrons window and the auger electrons window.


\subsection{Simulation results}

We used the numerical model described above to simulate the position measurements, which we  present below,  followed by the results of the polarization correction. To validate the method described above, we now describe the  result for the scanning incident positions.


\begin{figure*}[tb]
    \centering
    \includegraphics[width=\linewidth]{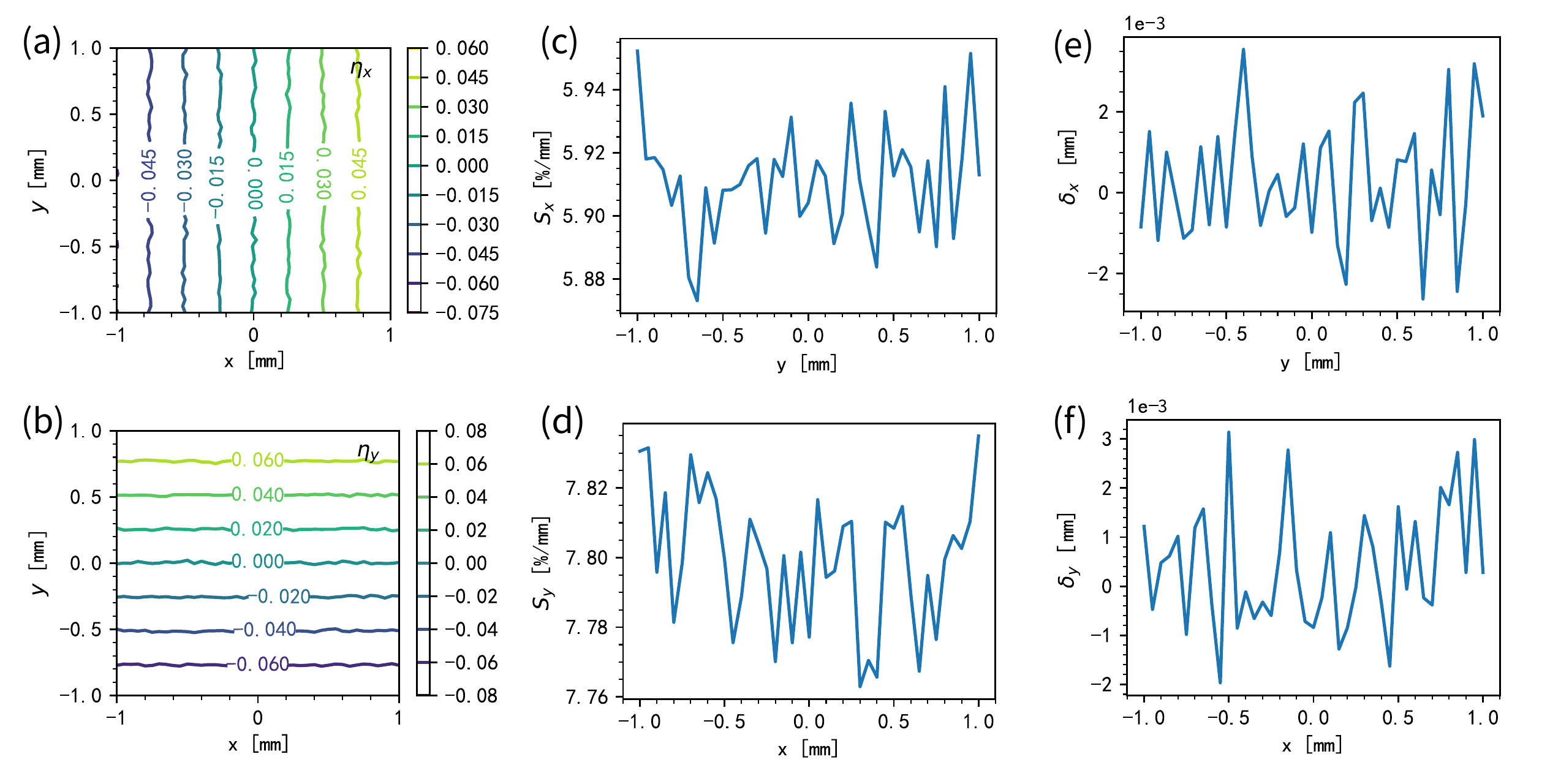}
    \caption{Results of simulation  of  transverse position measurement.  (a) Spatial distribution of \(\eta_x\). (b) Spatial distribution of \(\eta_y\). (c)  \(S_x\) as a function of   incident position \(y\).  (d)  \(S_y\) as a function of  incident position \(x\). (e)  \(\delta_x\) as a function of \(y\).  (f)  \(\delta_y\) as a function of \(x\).}
    \label{fig:fig1}
\end{figure*}

\begin{figure}[tbhp]
    \centering
    \includegraphics[width=\linewidth]{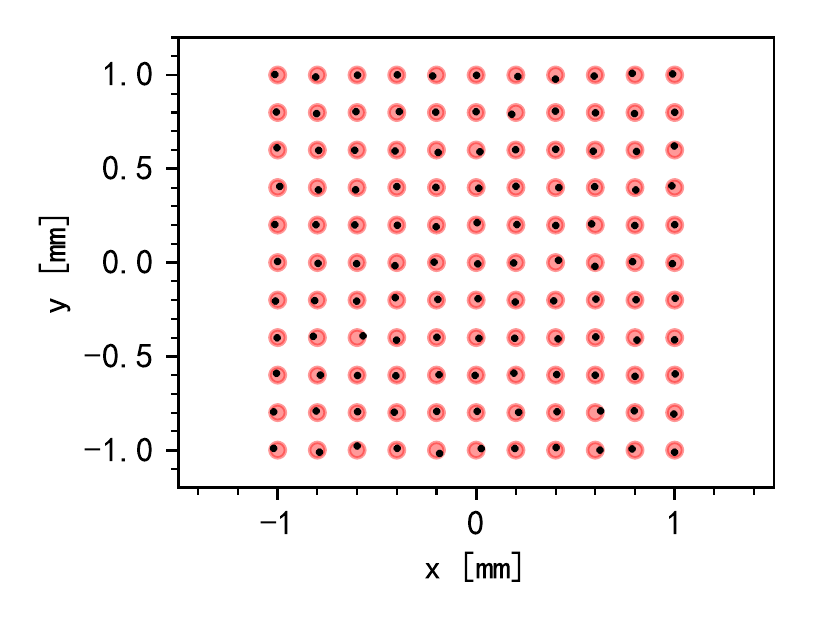}
    \caption{Circles show the real x-ray positions and  dots show the results of \(x=(1/S_x)\eta_x+\delta_x\), \(y=(1/S_y)\eta_y+\delta_y\) with the \(S_x=5.9\) [\%/mm], \(S_y=7.8\) [\%/mm], \(\delta_x=0.003\) [mm], \(\delta_y=0.004\) [mm].}
    \label{fig:xpos-meas}
\end{figure}

Figures \ref{fig:fig1}(a) and  \ref{fig:fig1}(b) show the  spatial distributions of \(\eta_x\) and \(\eta_y\). The ratio \(\eta_x\) increases linearly from \(-0.06\) to 0.06 as \(x\) goes from \(-1\)  to 1 mm, and the correlation between  \(\eta_x\) and  \(y\)  is sufficiently small  to be ignored.   \(\eta_y\)  increases linearly from \(-0.08\) to 0.08 as \(y\) goes from \(-1\)  to 1 mm, and the small correlation between \(\eta_y\) and  \(x\)  can also be ignored. The linearity can also be reported in terms of the gentle variation of \(S_x\) (or \(S_y\)). The sensitivity of \(\eta\)  remains constant as \(x\) (or \(y\))  varies within the region  \(y\ \in\) [\(-1\) mm, 1 mm] (or \(x\ \in\) [\(-1\) mm, 1 mm]), as shown in Figs. \ref{fig:fig1}(c) and  \ref{fig:fig1}(d). These results also show that \(S_x\) and \(S_y\) are about 5.9 [\%/mm] and 7.8 [\%/mm], respectively.  Figures \ref{fig:fig1}(e) and  \ref{fig:fig1}(f) allow us to conclude that the absolute values of \(\delta_x\) and \(\delta_y\) are less than \(3\times10^{-3}\)  and \(4\times10^{-3}\) mm, respectively. These figures also show that  \(\delta_x\) (\(\delta_y\)) is almost independent of \(y\)  (\(x\)), indicating that there is no need for a complicated position-dependent calibration. The values of \(S_x\) (\(S_y\)) and \(\delta x\) (\(\delta y\)) are  the average value along the \(y\)  (\(x\)) axis.  The unit vectors \(\hat{x}\) and \(\hat{y}\) were then calculated to predict position  as per Eqs. (\ref{eq:x predected}) and (\ref{eq:y predected}), with the \(S_x=5.9\) [\%/mm], \(S_y=7.8\) [\%/mm], \(\delta_x=0.003\) [mm], \(\delta_y=0.004\) [mm], and \(\psi=180^\circ\).  Figure \ref{fig:xpos-meas} shows the absolute deviation of the predicted position (\(x,y\)), in which \(x,y\in[-1\ \mathrm{mm},1\ \mathrm{mm}]\) are selected. The circles indicate the real x-ray position and the dots are calculated using Eq. (\ref{local coordinate to global coordinate}).

\begin{figure}[tb]
    \centering
        \includegraphics[width=\linewidth]{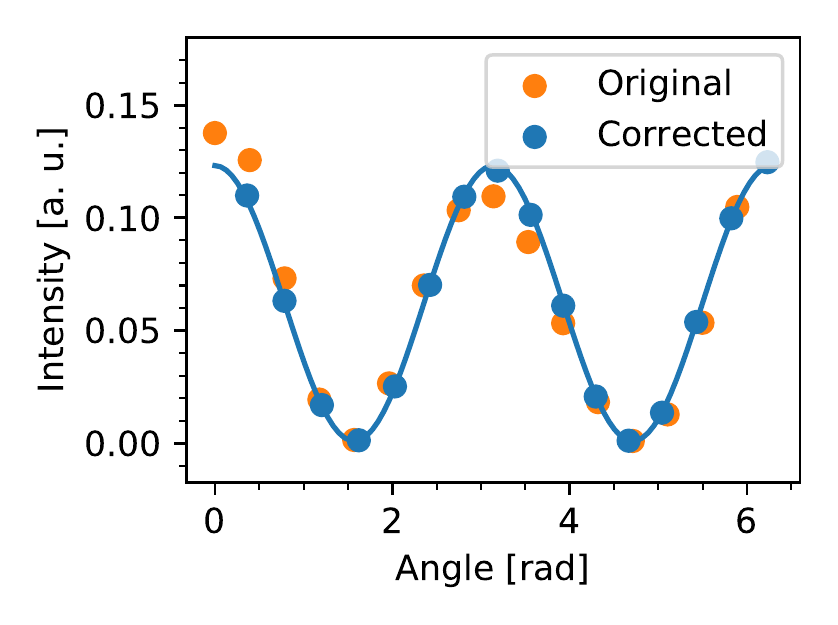}
        \caption{Angular distribution of photoelectrons before and after compensation.}
        \label{fig:corrected angular distribution}
\end{figure}

\begin{figure}[tb]
    \centering
    \includegraphics[width=\linewidth]{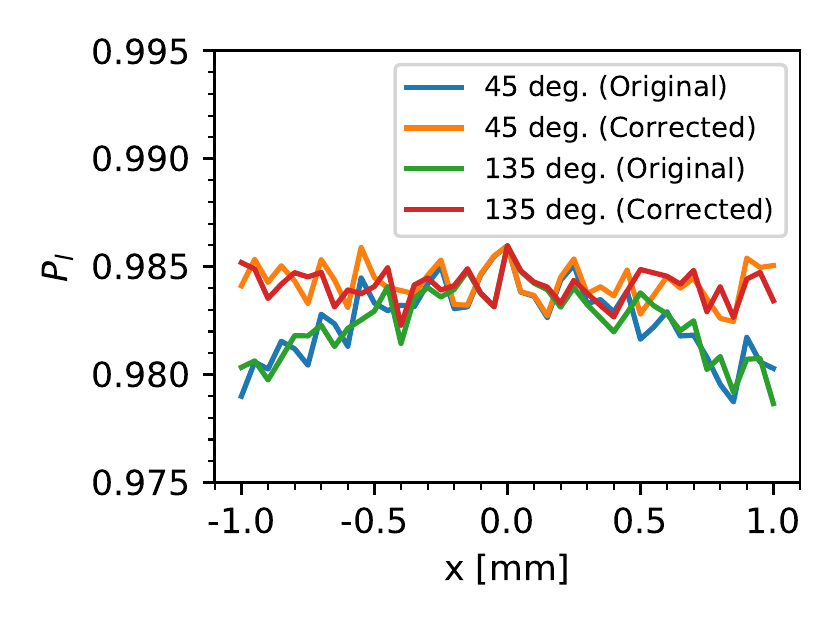}
    \caption{Original and compensated linear polarity along 45\(^\circ\) and 135\(^\circ\)  for horizontally polarized XFEL pulses.}
    \label{fig:polarization decompesation error}
\end{figure}

The intensity and \(\gamma\) angle of each detector depends strongly on the polarity and the incident position.
By correcting the incident position using Eqs. (\ref{eq:Angular distribution of phot}) and (\ref{eq:charge distribution at incident pos}), the angular distribution  \(p_s(\gamma)\) of photoelectrons is corrected and the FEL's polarization measurement  improves. 
Figure \ref{fig:corrected angular distribution} shows the original and corrected photoelectron angular distributions at the incident point \((x,y)=(1,1)\ \mathrm{mm}\).
Fitting with  Eq. (\ref{eq:Angular distribution of phot}) reveals a 40\% enhancement of the linear polarity  (from 0.979 to 0.985). Figure \ref{fig:polarization decompesation error} shows the linear polarity of XFEL pulses  along the axes oriented at 45\(^\circ\)  and 135\(^\circ\). These results show that the compensated value of \(P_l\) at all positions is close to \(P_l\)  at  \((x,y)=(0,0)\)~mm, which indicates that the transverse position-compensation method may be used for pulse-to-pulse polarization diagnosis.

\section{Calibration method}

\begin{figure}[tb]
    \centering
    \includegraphics[width=\linewidth]{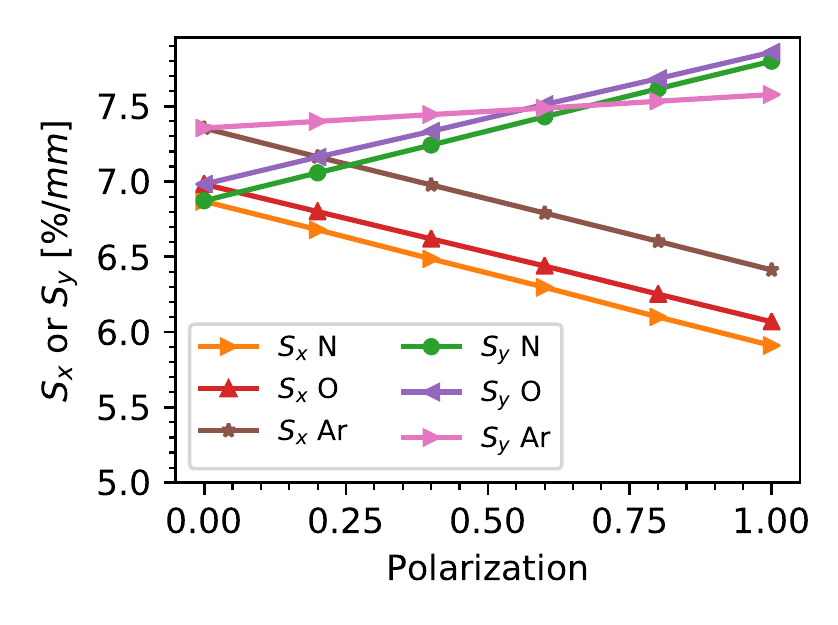}
    \caption{ \(S\) parameters of N, O, Ar as a function of linear polarization of FEL.}
    \label{fig:s-params-polar}
\end{figure}

Conventional XFELs are fully horizontally and linearly polarized. Thus, the value of \(S_x\) and \(S_y\) can be precisely calibrated by using an external XBPM such as fluorescent screens or backscattering detectors. But it is challenging and time consuming to calibrate  \(S_{x}\) (or \(S_y\))  for  polarization-modulated FEL pulses, which can be generated  using an additional elliptically polarized undulator  \cite{Deng2014design}.
When the linear polarity \(P_l\) of an x-ray pulse varies, the variations in \(S_x\) and \(S_y\) are caused by the variations in the  dipole-symmetric angular distribution of the photoelectrons. For a \(\text{N}_2\) target, Fig. \ref{fig:s-params-polar} shows the relationship between   \(S_x\) or \(S_y\) and the linear polarity, which suggests that \(S_x\) and \(S_y\) are linearly related to  \(P_l\) . In particular, \(S_x=S_y = S_0\) when the FEL pulses are fully circularly polarized because the angular distribution of photoinduced electrons is isotropic. 
As for the \(\text{N}_2\) target, the photoionization cross section of the \(p\) shell is significantly smaller than that of the \(s\) shell \cite{Band1979design,Trzhaskovskaya2002design}. It is reasonable to describe the linear relationship between the \(S\) parameters and the linear polarity \(P_l\) as
\begin{eqnarray}\label{eq:SParamLinearity}
    S_x &=& S_0+k_s P_l, \\
    S_y &=& S_0-k_s P_l,
\end{eqnarray}
where \(k_s = ({S_{x,1}-S_0})/{d_{P_l}}\), and
\begin{equation}\label{eq:S0}
    S_0 = \frac12[S_{x}(P_l=1)+S_{y}(P_l=1)].
\end{equation}
From the analysis above, the following approach can be applied to calibrate \(S_x\) and \(S_y\) for the \(\text{N}_2\) target: First, the sensitive values \(S_{x,1}\) and \(S_{y,1}\) for the fully linearly polarized FEL  are calibrated and \(S_0\) is obtained by using Eq. (\ref{eq:S0}). Next, the linear polarity without transverse position corrections \(P_{l,u}\) are measured and sensitivity values \(S_{x,1}\) and \(S_{y,1}\) are linearly interpolated using \(P_{l,u}\)  and  Eq. (\ref{eq:SParamLinearity}). Finally, the transverse position and the corrected polarization can be obtained by the compensation method described herein. 

For the \(\text{Ar}\) target, the photoionization cross sections for the \(p\) shells are of the same order of magnitude as those of the \(s\) shells, and Eqs. (\ref{eq:SParamLinearity}) and (\ref{eq:S0}) are no longer applicable. Therefore, the sensitivities \(S_{x,0}\) and \(S_{y,0}\) should be calibrated additionally for a circular polarized FEL to obtain \(k_s\) despite  electrons from the \(p\) shells being filtered by the electron-optical system.

\section{Discussion and Conclusion}

Although the method presented in this work is fully feasible, there remain challenges and further improvements to be made, which we discuss below.

The linearity and the stability of the charge measurement strongly affect  position monitoring and polarization diagnosis. The  charge detected in a channel is given as
\begin{equation}
    Q_i = n_e e \eta_{c,i} \eta_{p,i} G_i,
\end{equation}
where \(n_e\) are the number of electrons captured in  detection channel \(i\), \(e<0\) is the fundamental charge of an electron, \(\eta_c\) is the collection efficiency of  detection channel \(i\), \(\eta_p\) is  due to the incident position, and \(G_i\) is the gain of the MCP detector of  channel \(i\).
Because the electron-optical system can be configured with a precise static-electric bias, the collection efficiencies \(\eta_{c,i}\) are the same for all channels  at the time of the measurement. Thus, the most significant factor for measuring charge  is the gain  \(G_i\) of the MCP detectors.
As the capacity of residual charge decreases, the gain of the MCP detectors may decrease during the commissioning period. However, it is laborious to detach the polarimeter from the beamline and adjust the high-voltage power supplies to calibrate and ensure gain consistency for 16 channels  using a well-known x-ray  or electron source. A practical approach to calibrating the gain of the MCP detector is to use the MCP response to a single electron so that \(n_e = 1\) and \(\eta_{p,i}=1\). Single electrons can be obtained by adjusting the focusing electrodes.

Another factor that affects the position measurement is the yield of photoelectrons. As the number of photoelectrons increases, the  statistical fluctuations increase, which suggests that a stronger FEL intensity and a high target density  are preferable. In the current design of the ARPolar instrument, a specially designed skimmer serves to concentrate the gas direction, and two 700 L/s turbo pumps produce a target density of  \(1.54 \times 10^{-11}\ \text{kg/m}^3\), as determined by a  simulation with the well-benchmarked Molflow+ code \cite{2009MolflowIntroduction}. Note that position monitoring for multiple pulses is acceptable for compensating polarization diagnosis because  drift between the FEL machine states is sufficiently slow to measure the mass center of FEL pulses.

Methods exist to potentially measure  position  using an ARPolar-like instrument. Because electron energy can be reduced by using grid electrodes while their intensity is maintained by using focusing electrodes in the detection channels, a  synchronized, fast-rise-time electric field in front of the MCP detector can be used to filter  electrons with different flight times, such as is caused by the different flight distance between the transverse position of XFEL pulses and the detector.
  

This work thus describes methods and principles for  single-shot transverse position measurements and compensation for diagnosing FEL polarization states. A comprehensive numerical model is presented to analyze the accuracy and stability of the method. Polarization compensation reduces the error in \(P_l\)  by 40\% compared with a direct diagnosis. 

\begin{acknowledgments}
        This work was funded by National Key Research and Development Program of China (Grant No. 2016YFA0401901), the National Natural Science Foundation of China (Grant No. 11775293), and the Ten Thousand Talents Program.
\end{acknowledgments}

\bibliography{aipsamp}

\end{document}